# Graphene-assisted metagrating: from coherent to angular-asymmetric control of absorption and reflection

**ENSIYE GHORBANPOUR, SAHAR BEHROOZINIA, AND ALI ABDOLALI** [*]

*Applied Electromagnetic Laboratory, School of Electrical Engineering, Iran University of Science and Technology, Tehran, 1684613114, Iran.*
*\*abdolali@iust.ac.ir*

**Abstract:** In this paper, we exploit the metagrating paradigm to achieve coherent control of absorption and reflection in a two-port device. Employing graphene ribbon as a tunable element allows us to, for the first time, realize a reconfigurable metagrating that integrates diversified coherent functionalities into a single planar structure. It is illustrated that the suggested design can behave as a coherent perfect absorber at multiple operating incident angles offered by its period-reconfigurability. Besides, our proposed metagrating is also capable of highly-efficient dynamic beam steering based on the coherent interaction of light with light through finely adjusting its chemical potentials, and its compatibility is then investigated to realize linear all-optical logic gates. Moreover, we use the idea of graphene-based metagrating to put forward an extremely asymmetrical device which exhibits high retroreflection upon illumination from one side and a strong absorption under excitation from the opposite side, revealing a great enhancement in design and fabrication simplicity compared to the previous works. The EM response of our suggested device can also be switched from a strongly asymmetric to a symmetric behavior by merely modulating the chemical potentials of its graphene ribbons.

## 1. Introduction

The advent of coherent perfect absorption proposed based on the original concept of a time-reversed laser leads to a new type of interferometric optical system [1,2]. Compared to the traditional absorbers, which are under the illumination of a single beam, coherent perfect absorbers (CPAs) rely on the interplay of two, and even more, incident lights, yielding total absorption or transparency, or modulation of them. So far, lots of research based on various materials and structures such as dielectric slabs, graphene-based structures, coupled resonators, and particularly metamaterials [3-14] has been carried out to exploit this fundamental principle and report the properties of CPAs.

Beyond the absorption modulation offered by the CPA mechanism, the coherent light-light interaction on the planar interfaces can go much further. In [15], the signal processing opportunities presented by the coherent interaction of light have been studied. CPA-inspired logic gates are also explored [16-19], which operate much faster and require lower input power than their nonlinear-based counterparts, providing great potential applications in ultrafast all-optical data processing, image recognition, and data analysis. Besides, the possibility of dynamic redirection of light with light by employing the constructive and instructive interference of two interfering beams through ultrathin metasurfaces is also investigated [20,21]. In [21], following the generalized Snell's law, the specular and anomalous reflections have been coherently controlled and switched between on and off state by changing the phase difference between two coherent beams using finely designed phase gradient metasurfaces, two-dimensional structures consisting of subwavelength particles with gradient phase shift along the metasurface. However, recently, it has been revealed that graded metasurfaces, obeying the generalized Snell's law, are fundamentally limited on their overall efficiency in the transformation of an impinging wavefront [22]. These efficiency-related challenges and also the high precision demands for the fabrication of fastly varying impedance profiles have been relaxed by introducing the concept of metagrating [23]. Metagrating is a periodic structure in which the local period selects a discrete set of diffraction orders and engineering the constituent scatterers in each period enables highly efficient routing of light to the any arbitrary direction (which is aligned in the direction of one of the higher order modes). Over the past few years, significant attention has been devoted to this emerging area, and a variety of metagrating-based proposals capable of various functionalities such as perfect wave manipulation in both reflection and refraction [23-26] and highly-efficient control of diffraction pattern [27,28] have been put forward.

In this work, we offer a novel route based on the metagrating concept to designing coherently controllable devices without limitations associated with gradient metasurfaces. For the first time, a multifunctional platform that integrates diversified coherent functionalities from coherent perfect absorption and beam deflection to all-optical logic gates into a single planar structure is presented, which has not been explored yet, despite the remarkable achievements in the development of interference-assisted EM devices. At the supercell level, the designed metagrating consists of three graphene ribbons of different controllable chemical potentials. The suggested structure can serve as an angle-tunable coherent perfect absorber, perfectly absorbs two coherent lights at multiple operating angles. Furthermore, it is demonstrated that our proposed metagrating can also function as a highly-efficient coherent beam deflector with the ability to coupling almost all the incident energy into one port. Meanwhile, the potential application of the designed architecture in achieving three all-optical logical operations is also explored. We also inspect that our idea of metagrating based on graphene ribbons can be employed to design a reconfigurable asymmetric planar structure in which a large retroreflection is observed when the metagrating is excited from one side and at the same time, a near-perfect absorption occurs under the excitation of the opposite side. Besides, the amount of absorption can be tuned by choosing the proper chemical potentials of its particles. Our demonstration shows a significant enhancement in fabrication simplicity compared with the previous works. [29-31].

## 2. Design method

we design with a reflective metagrating composed of a periodic array of scatterers supporting only two propagative diffraction modes. As the first step, the periodicity of the structure $P$ is set to

$$P = \lambda/(2\sin\theta_i) \tag{1}$$

where $\lambda$ indicates the operating wavelength and $\theta_i$ is the angle of incidence. Due to the periodicity we chose, there exist only two propagation reflection modes, $n=-1$ and $n=0$, which are aligned in the directions of $\theta_0 = \theta_i$ and $\theta_{-1} = -\theta_i$, respectively. Let us assume that we want to design the metagrating for the incident angle $\theta_i = 45°$; the periodicity, in this case, is set to $P = \lambda/\sqrt{2}$, according to Eq.(1). As can be seen from Fig. 1 (a), when the electromagnetic (EM) wave impinges upon the metagrating at the considered angle, it splits into two parts: retroreflection and specular reflection corresponding to the orders of $n=-1$ and $n=0$, respectively. On the other hand, under the illumination of the opposite-side ($\theta_i = -45°$) incidence, a similar occurrence is observed, as illustrated by Fig. 1.(b). Therefore, the device can be regarded as a two-port network (see Fig. 1(c)) and described by a complex scattering matrix S, which connects the obliquely reflected waves $b_1$ and $b_2$ (output beams) with the opposite-side incidence $a_1$ and $a_2$ (input beams) as

$$\begin{bmatrix} b_1 \\ b_2 \end{bmatrix} = \begin{bmatrix} S_{11} & S_{12} \\ S_{21} & S_{22} \end{bmatrix} \begin{bmatrix} a_1 \\ a_2 \end{bmatrix} \quad (2)$$

in which, $S_{11} = r_{-1}^- \exp(i\alpha_{-1}^-)$, $S_{12} = r_0^+ \exp(i\alpha_0^+)$, $S_{21} = r_0^- \exp(i\alpha_0^-)$, and $S_{22} = r_{-1}^+ \exp(i\alpha_{-1}^+)$ are the complex reflection coefficients of the corresponding diffraction modes. The subscript indicates the mode order $n$ and the superscript - (+) represents the left-(right-) side incidence. Owing to the reciprocity, the specular reflections from both sides are identical, i.e., $S_{12} = S_{21}$, while the left-side retroreflection $S_{11}$ can be, generally, different from the retroreflection excited from the right-side incidence $S_{22}$. These three independent complex reflection coefficients, which describe the output intensities determined by the properties of the system, can be manipulated by engineering the scatterers in each period. From Eq. (2), the output reflection intensity at each port can be written as:

$$|b_1|^2 = (r_{-1}^-)^2 |a_1|^2 + (r_0^+)^2 |a_2|^2 + 2r_0^+ r_{-1}^- |a_1||a_2|\cos\left(\alpha_{-1}^- - \alpha_0^+ + \delta\right) \quad (3)$$

$$|b_2|^2 = (r_0^-)^2 |a_1|^2 + (r_{-1}^+)^2 |a_2|^2 + 2r_{-1}^+ r_0^- |a_1||a_2|\cos\left(\alpha_0^- - \alpha_{-1}^+ + \delta\right) \quad (4)$$

where the *cos* term refers to the interference between two reflected beams in each channel. Both factors, *i.e.*, the relative phase difference between incident waves ($\delta$) and also scattering parameters, $\Delta\Phi_1 = \alpha_{-1}^- - \alpha_0^+$ and $\Delta\Phi_2 = \alpha_0^- - \alpha_{-1}^+$, affect the interference in each channel.

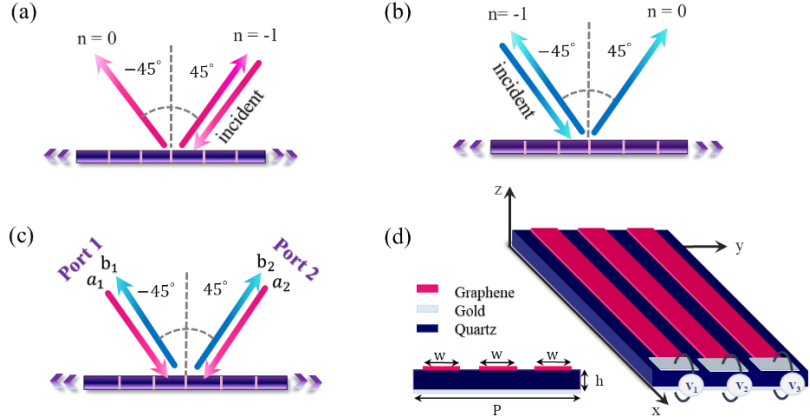

Fig. 1. Schematic view of (a) a reflective metagrating with two open floquet channels, for oblique incidence from $\theta_i = 45°$ only specular and retro reflection channels are open (b) Similar mechanism for the opposite incidence ($\theta_i = -45°$). (c) the metagrating as a two-port network, and (d) the designed supercell consisting of three graphene ribbons fed by different DC voltages.

As a demonstration purpose, we consider the topology of metagrating as shown in Fig. 1(d), consisting of three graphene ribbons placed on top of a grounded dielectric layer. All the graphene ribbons have the same widths, and their optical properties can be solely controlled using external biasing voltages. A supporting dielectric layer ($\varepsilon_r = 3.75$, loss tangent= 0.0004) with the thickness of $h$ is utilized as the core substrate, which is terminated by a gold layer with a thickness of 200 nm to avoid energy transmission. The EM properties of the gold material are characterized by the well-known Drude model [32]. The graphene ribbons have been considered as a 2D boundary with a complex surface conductivity $\sigma_s$ containing intraband conductivity $\sigma_s^{intra}$ and interband conductivity $\sigma_s^{inter}$, obeying the Kubo formula which is expressed below [33]:

$$\sigma_s(\omega, E_f, \Gamma, T) = \sigma_s^{intra}(\omega, E_f, \Gamma, T) + \sigma_s^{inter}(\omega, E_f, \Gamma, T) \quad (5)$$

$$\sigma_s^{intra}(\omega, E_f, \Gamma, T) = -j\frac{e^2 k_B T}{\pi \hbar^2 (\omega - j2\Gamma)}\left[\frac{E_f}{k_B T} + 2\ln\left(e^{\frac{-E_f}{k_B T}} + 1\right)\right] \quad (6)$$

$$\sigma_s^{inter}(\omega, E_f, \Gamma, T) = -j\frac{e^2}{4\pi \hbar}\left[\frac{2|E_f| - (\omega - j2\Gamma)\hbar}{2|E_f| + (\omega - j2\Gamma)\hbar}\right] \quad (7)$$

Here, $e$ is the electron charge, $E_f$ indicates the Fermi energy, $k_B$ is the Boltzmann's constant, $\hbar$ remarks the Plank constant, $\omega$ is the angular frequency, $T$ is the environment temperature and, $\Gamma = 1/2\tau$, $\tau$ represents the relaxation time. $T = 300K$ and $\tau = 1ps$ are considered throughout this paper. The surface conductivity of each graphene ribbon can be individually controlled using an external biasing voltage applied between the ground plane and the conductive pads placed on each graphene ribbon (see Fig. 1 (d)). The working frequency is chosen as f=5 THz. It should be pointed out that this number of meta-atoms, three graphene ribbons per period, can provide sufficient degrees of freedom to fully manipulate three independent complex scattering parameters at will by adjusting the current distribution on each ribbon [27,28].

### 3. coherent perfect absorption with the angle-tunability feature

As the first example, the metagrating is elaborately designed to serve as a coherent perfect absorber capable of totally absorbing two incident waves with the same amplitude and phase ($|a_1| = |a_2| = 1, \delta = 0$) coming from ports 1 and 2. In order to satisfy the coherent perfect absorption requirement, both output beams $b_1$ and $b_2$ should be suppressed. According to Eqs. (3), (4), it can be deduced that such conditions take place in the case of $S_{11} = -S_{12}$ and $S_{21} = -S_{22}$. Due to the reciprocity, these conditions lead to $S_{11} = S_{22}$, meaning that the device should pose a mirror symmetry, which can be achieved in our design by setting identical values for the chemical potential of the first and third graphene ribbons. It should be emphasized that to achieve CPA conditions, the presence of a lossy structure is essential so that realizing an ideal coherent perfect absorber with a maximum modulation rate of 100% requires 50% absorption under a single beam incident [34]. A single y-polarized EM plane wave excites the designed structure at Port 1, and its specular and anomalous reflections are measured numerically through a full-wave commercial software, CST Microwave Studio. The periodic boundary conditions were applied to the y-directed boundaries while Floquet ports were employed along the z-direction. A comprehensive parametric study has been accomplished to find the best structural parameters, including the widths of the graphene ribbons $w$, the thickness of the dielectric $h$, and the suitable chemical potentials $(\mu_c^1, \mu_c^2,$ and $\mu_c^3)$. The optimum values of parameters for this case are obtained as *w=11.6 μm, h=8.5 μm,* $\mu_c^1 = \mu_c^3 = 0.52$ *eV*, and $\mu_c^2 = 0.88$ *eV*. In Figs. 2 (a), (b) the amplitude and phase spectra of the scattering parameters, under the illumination of a single oblique beam at Port 1, are plotted. At the desired frequency, the scattering matrix of our two-port structure is:

$$S = \begin{bmatrix} 0.44\angle -113 & 0.47\angle 60 \\ 0.47\angle 60 & 0.44\angle -113 \end{bmatrix} \quad (8)$$

which can successfully meets the CPA requirements, i.e., $S_{11} = -S_{12}$ and $S_{21} = -S_{22}$. We define $o_1 = |b_1|^2/2$ and $o_2 = |b_2|^2/2$ as the normalized outgoing powers (respect to the total incoming power) at ports 1 and 2, and thus to determine the coherent absorptivity, we can use the

$$A_C = 1 - O_1 - O_2 \tag{9}$$

Based on Eq. (9), more than 99% of the coherent lights are absorbed by our simple structure at the frequency of 5 THz. As can be deduced from Eqs. (3), (4), the output power in each port can be modulated easily by adjusting the relative phase difference between the two incident light beams. Fig. 2(c) shows the coherent absorption and the normalized power in each output channel as a function of the phase difference between the input signals at 5 THz. The results inspect that in the case of zero relative phase difference, the absorptivity peaks at 0.99 while it reaches its lowest point at 0.18 for a π phase difference where 41% of total incident energy is coupled to each port and the metagrating behaves as a coherent beam splitter.

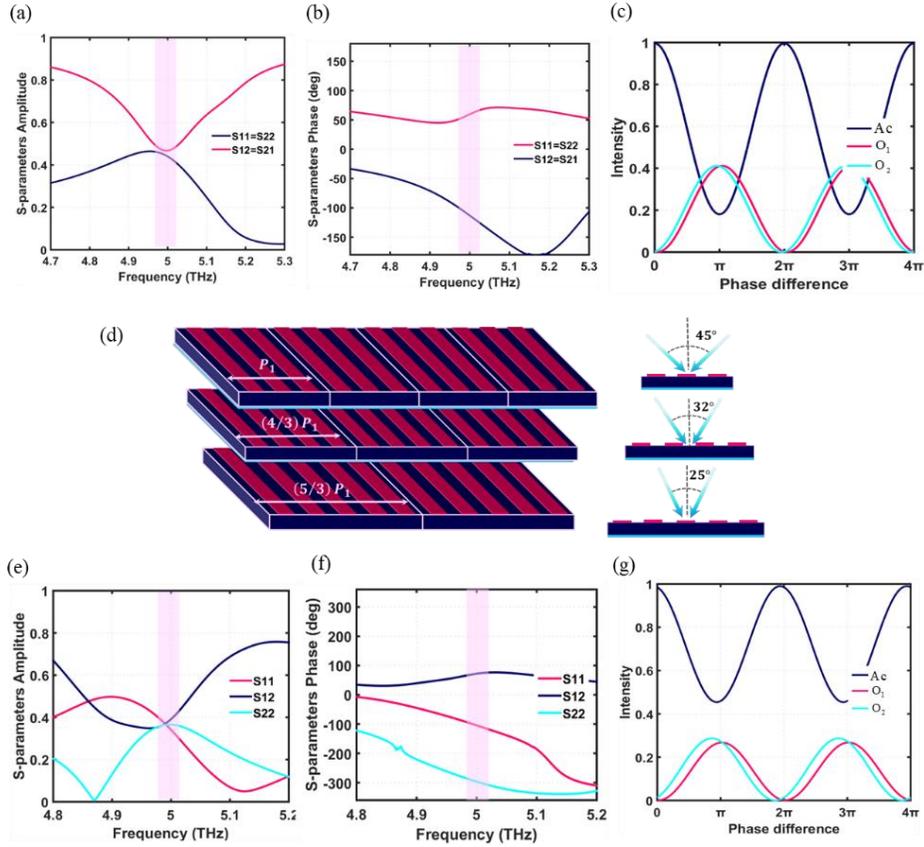

Fig. 2. Realizing an angle-tunable coherent perfect absorber. (a) The magnitude and (b) phase of the scattering parameters related to the metagrating acting as a CPA at $\theta_i = \pm 45°$ (the results under the excitation of ports 1 and 2 are the same), and (c) the phase-dependent outgoing powers and coherent absorption normalized by the total incoming power. (d) The periodicity tunability of the proposed metagrating-based CPA resulting in CPAs operating at different incident angles $\theta_i = \pm 45°$, $\theta_i = \pm 32°$, and $\theta_i = \pm 25.1°$, through systematically tuning the chemical potentials of its graphene ribbons. (e) The magnitude and (f) phase spectra of the reflections for the metagrating-based CPA operating at $\theta_i = \pm 32°$ (g) the phase-dependent outgoing powers and coherent absorption normalized by the total incoming power for CPA at $\theta_i = \pm 32°$.

It is worth mentioning that our suggested approach can also be established to design CPAs at multiple arbitrarily-selected incident angles ($\theta_i > 20°$) by finely selecting the periodicity of the structure, according to Eq. (1). It should be pointed out that the angles smaller than 20° are not permissible since, in this region, the metagrating cannot play the role of a two-port system with only two open Floquet modes. Interestingly, for the proposed CPA, once the structure is designed to operate at $\theta_1$ with the periodicity of $P_1$, the equal widths of graphene strips while being separately controlled through DC biasing voltages, allows one to switch the periodicity of metagrating between $P_1$, $4P_1/3$, $5P_1/3$, and so on. Indeed, the design can be dynamically changed so that the repetitive part of the structure is occupied with 3, 4, and 5 graphene ribbons, respectively, as plotted in Fig. 2(d).

The adjustability of the metagrating periodicity brings up a degree of freedom to engineer the angular direction of available channels. Given Eq. (1), the change in the periodicity of metagrating $P_{new} = \left(1 + \dfrac{m}{3}\right)P$, subject to the proper selection of chemical potentials, leads to new two-port CPA configurations, for which the channel orientations $\theta_m$ can be defined as:

$$\theta_{m+1} = \sin^{-1}(\lambda / 2 \left(1 + \dfrac{m}{3}\right) P_1) \ , \ m = 1, 2, 3, \ldots \qquad (10)$$

As can be included from Eq. (10), the yield operating angles are smaller than the first angle $\theta_1$ for which the metagrating has been primarily designed. For our designed CPA operating at $\theta_1 = 45°$, the other possible angles will obtain as $\theta_2 = 32°$, $\theta_3 = 25.1°$, and $\theta_4 = 20.7°$ for m=1, 2, and 3, respectively, based on Eq. (10). As a proof of concept, we found proper chemical potentials when the proposed CPA is aimed at playing the role of CPA at $\theta_2 = 32°$. Taking the same procedure as the first CPA realized in this paper, the best values for the chemical potentials of four contributing graphene ribbons obtained through applying a parametric optimization are *0.83 eV*, *0.55 eV*, *0.89 eV*, and *0 eV* for two-beam excitation at $\theta_i = \pm 32°$. The corresponding numerical results are plotted in Figs. 2 (e)-(g), indicating that at the desired frequency of 5 THz, the proposed structure fulfills the CPA requirements on the scattering parameters and can successfully operate as CPA excited by coherent in-phase beams with the incident angle of $\theta_i = \pm 32°$ from the left and right sides. Fig. 2(g) shows the phase-dependent behavior of the coherent absorption, remarking that as the input phase difference varies from 0 to π, the coherent absorption is modulated in the range of *0.42* to *0.99*.

## 4. coherent beam deflection and all-optical logical operations

In this section, we first consider the metagrating as a lossless beam splitter in which under the single beam incident from ports 1 and 2, the power splits equally with the power ratio 50:50 into output ports, meaning that $|S_{11}|^2 = |S_{21}|^2 = |S_{12}|^2 = |S_{22}|^2 = 0.5$. Now, assuming two coherent excitations with the same amplitude and relative phase ($|a_1| = |a_2| = 1, \delta = 0$), if the scattering parameters have the proper phase difference of $\Delta\Phi_1 = \pi$ and $\Delta\Phi_2 = 0$, a destructive interference in Port 1 and a constructive one in Port 2 will occur. In other words, all the incoming energy is channeled into Port 2 while no energy is coupled to the other one. However, when two incident lights with the relative phase of π excite the metagrating, the incident energy is fully transferred to Port 1, owing to the constructive and destructive interplay between scattered waves in Port 1 and 2, respectively. It should be pointed out that in this case, in contrast to the previous example, the left-and right-side retroreflections are not identical and should satisfy $S_{11} = -S_{22}$, which exhibit an asymmetric behavior requiring an unsymmetrical structure to fulfill the desire. The optimum group of chemical potentials for this case are obtained as *0.03 eV*, *0.6 eV*, and *1.4 eV*, respectively. As can be seen, the chemical potential of the third ribbon is different from the first one, exhibiting the asymmetrical nature of the designed metagrating. The widths of graphene ribbons and the thickness of the substrate are the same as presented for the previous examples. Figs. 3(a), (b) depict the amplitude and phase spectra of the scattering parameters under the illumination of a single oblique beam at ports 1 and 2. The corresponding two-port scattering matrix of the structure, at the designed frequency, reads:

$$S = \begin{bmatrix} 0.67\angle -138.4 & 0.64\angle 56.7 \\ 0.64\angle 56.7 & 0.68\angle 72.6 \end{bmatrix} \quad (11)$$

which can effectively satisfy the above-mentioned conditions for efficient coherent beam deflection, *i.e.*, $S_{11} = -S_{12}$ and $S_{21} = S_{22}$. As can be observed from Fig. 3(c), the power ratio between Port 1 and Port 2 can be elaborately modulated under the variation of the phase difference between the two incident beams so that at zero ($\pi$) phase difference for two coherent excitations about 87% of the total incoming power is efficiently coupled into the Port 2 (1) while almost no energy (1.5%) can be detected in the other port (about 11.5% of the incident energy is absorbed due to the ohmic loss of graphene), and thus our proposed device can efficiently play the role of all-optical switchable beam deflector.

In the following, we will demonstrate that such coherent control of reflected fields at each port provides a route for realizing all-optical linear logic gates. Table.1 describes three basic logical operations, *i.e.*, AND, OR, XOR. We can consider the two incident light beams impinging on the designed metagrating analogous to the two ports of logic gates, A and B (the relative phase difference between inputs is assumed to be 0). The resultant intensities at the output ports corresponding to different combinations of input channels are inspected in Figs. 3 (d)-(f). As can be observed, for a set of input channel intensities of (0,0), (0,1), (1,0), and (1,1), the output intensity at Ports 1 reaches 0, 0.4, 0.45 and 0.03, respectively. In parallel, the values of 0, 0.46, 0.4, and 1.74 are detected for the output intensity at Port 2 corresponding to the input logical states of (0,0), (0,1), (1,0), and (1,1), respectively. By introducing a proper threshold intensity through which the output status is determined as 0 or 1(values greater than the threshold is regarded as logical 1), one can define different all-optical logic operations for the left- and right-side output channels. By choosing the threshold intensity to be a value between 0 and 0.4, the designed metagrating can respectively play the roles of XOR and OR gates in ports 1 and 2, and at the same time, the AND operation is also implemented at Port 2 if the threshold is set to a value 0.46 to 1.74, as it is apparent from Figs. 3 (d)-(f). It is rewarding to note that, compared to the previous logic gates based on the CPA- mechanism suggested in the literature [21,23], where different operations require different relative phase between input signals, our proposal can perform all these three functions simultaneously without any need to change the input phase difference.

Table 1. The truth table of three basic logical operations, AND, XOR, and OR.

| Input | | Output | | |
|---|---|---|---|---|
| A | B | A XOR B | A OR B | A AND B |
| 0 | 0 | 0 | 0 | 0 |
| 1 | 0 | 1 | 1 | 0 |
| 0 | 1 | 1 | 1 | 0 |
| 1 | 1 | 0 | 1 | 1 |

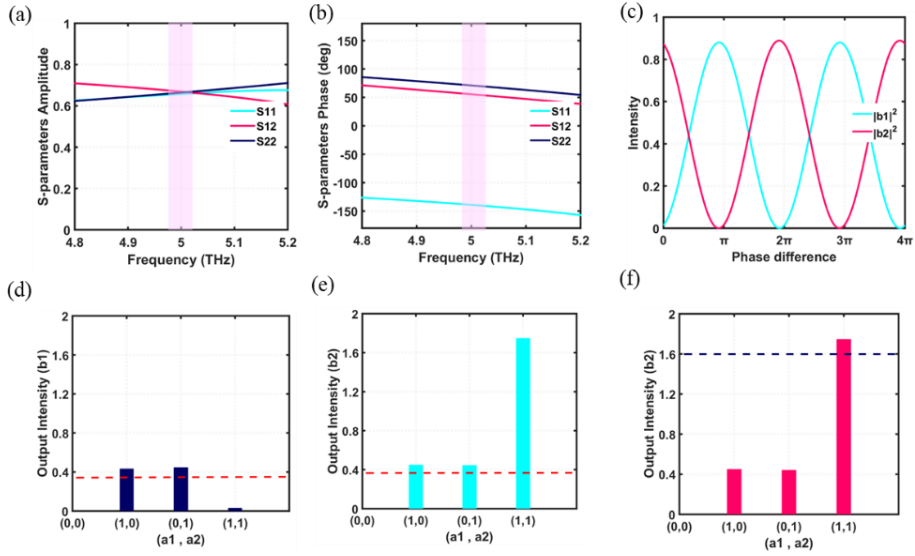

Fig. 3. All-optical beam deflection and linear logic gates. (a), (b) The magnitude, and phase spectra of the reflections under single beam incidence from both port 1 and 2. (c) Relative output powers (normalized to the total incoming power) as a function of the relative phase difference at the selected frequency of 5 THz. (d), (e) and (f) output intensities for different combinations of inputs for XOR, OR, and AND logic gates, respectively.

## 5. Reconfigurable angular-asymmetric absorption

As mentioned above, our proposed metagrating in the previous section reveals an asymmetric response having the same retroreflection amplitude but different phases. This angular asymmetry could be extended to the case where the scattered energy levels upon illuminating from two oppositely-side incidences are different. This feature is of considerable importance in practical applications where directional responses are required. A particular scenario of asymmetric wave manipulation occurs when all incident energy is perfectly reflected back to the source direction upon illumination from one side and completely absorbed when excited from the other side [29-31]. It may be well to note that, although it is relatively easy to have a zero reflection from one side, it is strongly challenging to have, at the same time, high reflectivity from the opposite side. In the following, we will show that our proposed metagrating-based approach can also be applied to achieve such asymmetric response through a simple planar structure.

We explore a new structure similar to the previous design but including graphene ribbons with different (see the inset of Fig. 4(a)). We seek the six contributing parameters, *i.e.,* three chemical potentials and three widths of ribbons $w_1$, $w_2$, and $w_3$, in order to realize a metagrating having highly-efficient retroreflection when the wave comes from one side ($\theta_i = -45°$) and near-perfect absorption when the wave comes from the opposite side ($\theta_i = 45°$). The best parameters obtained from the optimization process are *1.33 eV, 1.46 eV, 0.66 eV,* and *5.2 μm, 8.8 μm, 3.8 μm*, which respectively represent the chemical potentials and the width of the graphene ribbons. The reflection spectra and the y-directed component of the scattered electric field for both incident angles of $\theta_i = -45°$ (Port 1) and $\theta_i = 45°$ (Port 2) are depicted in Figs. 4(a) and (b), respectively. The numerical results indicate that our proposed architecture can effectively absorb 94% of the incident power coming from $\theta_i = 45°$ and realize retro-reflection with high efficiency of 74% when excited from the opposite direction. It is worth to emphasize that our design based on the metagrating principals with only three subunits relaxes the fabrication and design complexities compared to the previous proposals based on the surface impedance approach where more than 10 complex subunits should be included in one period [30,31]. Furthermore, if the chemical potentials of the first and second ribbons keep fixed and the third one decreases from *0.66 eV* to *0.55 eV*, the retroreflection efficiency at the left-side port stays high at 0.74 (see Fig. 4. (c)) while the efficiency of right-sided retroreflection increases from 0.05 to 0.74 (the specular reflection efficiency also remains lower than 0.01), which means that the absorption of light incident from Port 2 can be modulated from 94% to 25%, as can be seen from Fig.4(d). In other words, our planar structure can be switched between a strongly asymmetric and a symmetric state simply by applying appropriate external voltages, which to the best of the authors' knowledge, has not been explored before.

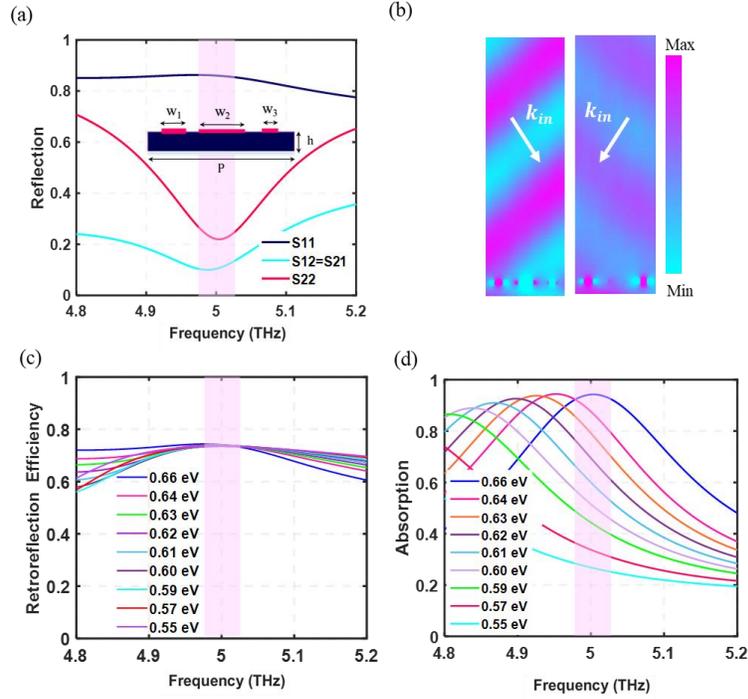

Fig. 4. Reconfigurable angular-asymmetric absorber. (a) The specular reflection and retroreflection from both incident angles of $\theta_i = -45°$ and $\theta_i = 45°$. The front view of our proposed asymmetric unitcell constructed by three graphene ribbons with different widths is illustrated in the inset of the figure. (b) the y-component of the scattered electric field for two scenarios of $\theta_i = -45°$ and $\theta_i = 45°$, respectively. (c) the frequency spectra of retroreflection efficiency under the illumination of $\theta_i = -45°$ and (d) absorption upon excitation from $\theta_i = 45°$ for the proposed design in which $\mu_c^1$ and $\mu_c^2$ are respectively fixed at *1.33 eV* and *1.46 eV*, while $\mu_c^3$ varies from *0.66 eV* to *0.55 eV*.

## 6. Conclusion

In summary, a novel approach to design coherent devices based on the metagrating concept has been introduced. Our reconfigurable two-port metagrating constructed of three graphene ribbons has been explored to realize diverse coherent scattering functionalities. The concept of angle-tunable coherent perfect absorption, the ability to act as a CPA in different operation angles for coherent input beams, provided by period-reconfigurability of our proposed structure, has been introduced. We numerically demonstrated that our suggested structure can also operate as a highly efficient all-optical beam deflector where almost the whole incident energy has been channeled into one port while the output power leaving the other port vanishes. Besides, its potential was further explored to achieve linear logic gates, XOR, AND, and OR. Our proposed approach could be further developed in realizing complex coherently controllable devices having more than two input ports, exposing a variety of functionalities. We also inspected that our metagrating-based idea can also be employed to design an extremely asymmetric device having a high retroreflection under illumination from one side and strong absorption from the opposite side with the added benefit of absorption-tunability, through a simple planar structure.

**References**

1. Chong, Y. D., Li Ge, Hui Cao, and A. Douglas Stone. "Coherent perfect absorbers: time-reversed lasers." Physical review letters 105, no. 5 (2010): 053901.
2. Longhi, Stefano. "Coherent perfect absorption in a homogeneously broadened two-level medium." Physical Review A 83, no. 5 (2011): 055804.
3. Wan, Wenjie, Yidong Chong, Li Ge, Heeso Noh, A. Douglas Stone, and Hui Cao. "Time-reversed lasing and interferometric control of absorption." Science 331, no. 6019 (2011): 889-892.
4. Pirruccio, Giuseppe, Luis Martin Moreno, Gabriel Lozano, and Jaime Gómez Rivas. "Coherent and broadband enhanced optical absorption in graphene." ACS nano 7, no. 6 (2013): 4810-4817.
5. Fan, Yuancheng, Fuli Zhang, Qian Zhao, Zeyong Wei, and Hongqiang Li. "Tunable terahertz coherent perfect absorption in a monolayer graphene." Optics letters 39, no. 21 (2014): 6269-6272.
6. Ning, Yaying, Zhewei Dong, Jiangnan Si, and Xiaoxu Deng. "Tunable polarization-independent coherent perfect absorber based on a metal-graphene nanostructure." Optics Express 25, no. 26 (2017): 32467-32474.
7. Feng, Xiong, Jinglan Zou, Wei Xu, Zhihong Zhu, Xiaodong Yuan, Jianfa Zhang, and Shiqiao Qin. "Coherent perfect absorption and asymmetric interferometric light-light control in graphene with resonant dielectric nanostructures." Optics Express 26, no. 22 (2018): 29183-29191.
8. Sun, Yong, Wei Tan, Hong-qiang Li, Jensen Li, and Hong Chen. "Experimental demonstration of a coherent perfect absorber with PT phase transition." Physical review letters 112, no. 14 (2014): 143903.
9. Baranov, Denis G., Alex Krasnok, Timur Shegai, Andrea Alù, and Yidong Chong. "Coherent perfect absorbers: linear control of light with light." Nature Reviews Materials 2, no. 12 (2017): 1-14.
10. Zhang, Jianfa, Kevin F. MacDonald, and Nikolay I. Zheludev. "Controlling light-with-light without nonlinearity." Light: Science & Applications 1, no. 7 (2012): e18-e18.
11. Kang, Ming, Fu Liu, Teng-Fei Li, Qing-Hua Guo, Jensen Li, and Jing Chen. "Polarization-independent coherent perfect absorption by a dipole-like metasurface." Optics letters 38, no. 16 (2013): 3086-3088.
12. Zhu, Weiren, Fajun Xiao, Ming Kang, and Malin Premaratne. "Coherent perfect absorption in an all-dielectric metasurface." Applied Physics Letters 108, no. 12 (2016): 121901.
13. Nie, Guangyu, Quanchao Shi, Zheng Zhu, and Jinhui Shi. "Selective coherent perfect absorption in metamaterials." Applied Physics Letters 105, no. 20 (2014): 201909.
14. Fang, Xu, Kevin F. MacDonald, Eric Plum, and Nikolay I. Zheludev. "Coherent control of light-matter interactions in polarization standing waves." Scientific reports 6, no. 1 (2016): 1-7.
15. Fang, Xu, Kevin F. MacDonald, and Nikolay I. Zheludev. "Controlling light with light using coherent metadevices: all-optical transistor, summator and invertor." Light: Science & Applications 4, no. 5 (2015): e292-e292.
16. Papaioannou, Maria, Eric Plum, João Valente, Edward TF Rogers, and Nikolay I. Zheludev. "Two-dimensional control of light with light on metasurfaces." Light: Science & Applications 5, no. 4 (2016): e16070-e16070.
17. Papaioannou, Maria, Eric Plum, Joao Valente, Edward TF Rogers, and Nikolay I. Zheludev. "All-optical multichannel logic based on coherent perfect absorption in a plasmonic metamaterial." APL Photonics 1, no. 9 (2016): 090801.
18. Xomalis, Angelos, Iosif Demirtzioglou, Eric Plum, Yongmin Jung, Venkatram Nalla, Cosimo Lacava, Kevin F. MacDonald, Periklis Petropoulos, David J. Richardson, and Nikolay I. Zheludev. "Fibre-optic metadevice for all-optical signal modulation based on coherent absorption." Nature communications 9, no. 1 (2018): 1-7.
19. Meymand, Roya Ebrahimi, Ali Soleymani, and Nosrat Granpayeh. "All-optical AND, OR, and XOR logic gates based on coherent perfect absorption in graphene-based metasurface at terahertz region." Optics Communications 458 (2020): 124772.
20. Shi, Jinhui, Xu Fang, Edward TF Rogers, Eric Plum, Kevin F. MacDonald, and Nikolay I. Zheludev. "Coherent control of Snell's law at metasurfaces." Optics express 22, no. 17 (2014): 21051-21060.
21. Kita, Shota, Kenta Takata, Masaaki Ono, Kengo Nozaki, Eiichi Kuramochi, Koji Takeda, and Masaya Notomi. "Coherent control of high efficiency metasurface beam deflectors with a back partial reflector." Apl Photonics 2, no. 4 (2017): 046104.

22. Estakhri, Nasim Mohammadi, and Andrea Alù. "Wave-front transformation with gradient metasurfaces." Physical Review X 6, no. 4 (2016): 041008.
23. Ra'di, Younes, Dimitrios L. Sounas, and Andrea Alù. "Metagratings: Beyond the limits of graded metasurfaces for wave front control." Physical review letters 119, no. 6 (2017): 067404.
24. Chalabi, H., Y. Ra'Di, D. L. Sounas, and A. Alù. "Efficient anomalous reflection through near-field interactions in metasurfaces." Physical Review B 96, no. 7 (2017): 075432.
25. "Unveiling the properties of metagratings via a detailed analytical model for synthesis and analysis." Physical Review Applied 8, no. 5 (2017): 054037.
26. Epstein, Ariel, and Oshri Rabinovich. "Perfect anomalous refraction with metagratings." 12th European Conference on Antennas and Propagation (EuCAP 2018). IET, 2018.
27. Popov, Vladislav, Fabrice Boust, and Shah Nawaz Burokur. "Controlling diffraction patterns with metagratings." Physical Review Applied 10, no. 1 (2018): 011002.
28. Behroozinia, Sahar, Hamid Rajabalipanah, and Ali Abdolali. "Real-time terahertz wave channeling via multifunctional metagratings: a sparse array of all-graphene scatterers." Optics Letters 45, no. 4 (2020): 795-798.
29. Wang, Xu-Chen, Ana Díaz-Rubio, Viktar S. Asadchy, and Sergei A. Tretyakov. "Concept of an asymmetric metasurface absorber." (2018): 299-3.
30. Wang, Xuchen, Ana Díaz-Rubio, Viktar S. Asadchy, Grigorii Ptitcyn, Andrey A. Generalov, Juha Ala Laurinaho, and Sergei A. Tretyakov. "Extreme asymmetry in metasurfaces via evanescent fields engineering: Angular-asymmetric absorption." Physical Review Letters 121, no. 25 (2018): 256802.
31. Cao, Yanyan, Yangyang Fu, Qingjia Zhou, Xin Ou, Lei Gao, Huanyang Chen, and Yadong Xu. "Mechanism behind angularly asymmetric diffraction in phase-gradient metasurfaces." Physical Review Applied 12, no. 2 (2019): 024006.
32. Walther, M., D. G. Cooke, C. Sherstan, M. Hajar, M. R. Freeman, and F. A. Hegmann. "Terahertz conductivity of thin gold films at the metal-insulator percolation transition." Physical Review B 76, no. 12 (2007): 125408.
33. Chern, Ruey-Lin, and Dezhuan Han. "Nonlocal optical properties in periodic lattice of graphene layers." Optics express 22.4 (2014): 4817-4829.
34. Jeffers, John. "Interference and the lossless lossy beam splitter." Journal of Modern Optics 47, no. 11 (2000): 1819-1824.